\documentclass[doublespacing]{elsart}

\usepackage{amssymb}
\usepackage{graphicx}
\usepackage{amsmath}

\def\beq{\begin{equation}}
\def\eeq{\end{equation}}
\def\rmd{{\rm d}}

\journal{Physics Letters A}

\begin{document}

\begin{frontmatter}

\title{Neutrino oscillations in the field of a rotating deformed mass}

\author[fisica]{A. Geralico},
\ead{geralico@icra.it}
\author[fisica,messico]{O. Luongo}
\ead{orlando.luongo@roma1.infn.it}

\address[fisica]
{Physics Department and ICRA, University of Rome ``La Sapienza,'' I-00185 Rome, Italy}

\address[messico]
{Institute of Nuclear Science, University of Mexico, Mexico}

\begin{abstract}
The neutrino oscillations in the field of a rotating deformed mass is investigated.
The phase shift is evaluated in the case of weak field limit, slow rotation and small deformation.
To this aim the Hartle-Thorne metric is used, which is an approximate solution of the vacuum Einstein equations  accurate to second order in the rotation parameter $a/M$ and to first order in the mass quadrupole moment $q$.
Implications on atmospheric, solar and astrophysical neutrinos are discussed.
\end{abstract}

\begin{keyword}
Neutrino oscillation \sep Hartle-Thorne metric
\PACS 04.20.Cv 
\end{keyword}

\end{frontmatter}

\section{Introduction}

In the Standard Model with minimal particle content neutrinos are massless left-handed fermions.
The question whether neutrinos have a non-vanishing rest mass influences research areas from particle physics up to cosmology, but it remains an open issue \cite{report}. 
At present all hints for neutrino masses are connected with neutrino oscillation effects, namely the solar neutrino deficit, the atmospheric neutrino anomaly and the evidence from the LSND experiment \cite{aguilar}.
Possible extensions of the Standard Model to generate neutrino masses are reviewed, e.g., in Ref. \cite{zuber}.

Mass neutrino mixing and oscillation in flat spacetime were proposed by Pontecorvo \cite{pontecorvo}.
Later on Mikheyev, Smirnov and Wolfenstein \cite{msw} investigated the effect of transformation of one neutrino
flavor into another in a medium with varying density.
There have been many experimental studies exploring the evidence for oscillations of both atmospheric and solar neutrinos as well as imposing limits on their masses and mixing angle (see, e.g., Ref. \cite{esperimenti} and references therein).

The possibility to detect CP violation effects in neutrino oscillations by future experiments has also been considered in recent years \cite{klinkhamer,schwetz,gava,altarelli}.  
Neutrino oscillation experiments are expected to provide stringent bounds on many quantum gravity models entailing violation of Lorentz invariance, so allowing to test quantum gravity theories \cite{mavromatos1,mavromatos2}.
Planck scale-induced deviations from the standard oscillation length may be observable for ultra-high-energy neutrinos emitted by galactic and extragalactic sources by means of the next generation neutrino detectors such as IceCube and ANITA \cite{christian}.
Furthermore, since neutrinos can propagate freely over large distances and can therefore pile up minimal length effects beyond detectable thresholds, there is the possibility to explore the presence of a quantum-gravity-induced minimal length using neutrino oscillation probabilities \cite{sprenger}.

The effect of gravitation on the neutrino oscillations has been extensively investigated in the recent literature, starting from the pionering work of Stodolsky \cite{stodolsky}.
The correction to the phase difference of neutrino mass eigenstates due to the spherically symmetric gravitational field described by the Schwarzschild metric was calculated in various papers within the WKB approximation \cite{schw1,schw2,schw3,schw4,schw5}. The results obtained in these papers differ from each other due to different methods used to perform the calculation.
For instance, calculating the phase along the timelike geodesic line will produce a factor of 2 in the high energy limit, compared with the value along the null line \cite{nullvstime}.
A different method was proposed by Linet and Teyssandier \cite{linet}, based on the world function developed by Synge \cite{synge} and defined as half the square of the spacetime distance between two generic points connected by a geodesic path.
Unfortunately, the calculation of the world function is not a trivial task. In general, it is performed perturbatively unless the solution of the geodesic equations is explicitly known, as in the very special cases of Minkowski, G\"odel, de Sitter spacetimes and the metric of a homogeneous gravitational field \cite{gps}.
The effect of spacetime rotation on neutrino oscillations has been investigated in Ref. \cite{kerr}, where the Kerr solution was considered.
A mechanism to generate pulsar kicks based on the spin flavor conversion of neutrinos propagating in a slowly rotating Kerr spacetime described by the Lense-Thirring metric has been recently proposed \cite{lambiase1}.
Furthermore, the neutrino geometrical optics in a gravitational field and in particular in a Lense-Thirring background has been investigated \cite{lambiase2}. 
Finally, in Ref. \cite{KN} the generalization to the case of a Kerr-Newman spacetime has been discussed.

In the present paper we calculate the phase shift in the gravitational field produced by a massive, slowly rotating and quasi-spherical object, described by the Hartle-Thorne metric.
This is an approximate solution of the vacuum Einstein equations accurate to second order in the rotation parameter $a/M$ and to first order in the mass quadrupole moment $q$, generalizing the Lense-Thirring metric.
We then discuss possible implications on atmospheric, solar and astrophysical neutrinos.
The units $G=c=\hbar=1$ are used throughout the paper.

\section{Stationary axisymmetric spacetimes and neutrino oscillation}

The line element corresponding to a general stationary axisymmetric solution of the vacuum Einstein equations can be written in the Weyl-Lewis-Papapetrou \cite{weyl17,lew32,pap66} form as 
\begin{eqnarray}
\label{metgen}
\rmd s^2&=&-f(\rmd t-\omega \rmd\phi)^2\nonumber\\
&&+\frac{\sigma^2}{f}\left\{e^{2\gamma}\left(x^2-y^2\right)\left(\frac{\rmd x^2}{x^2-1}+\frac{\rmd y^2}{1-y^2}\right)+(x^2-1)(1-y^2)\rmd\phi^2\right\}\ 
\end{eqnarray}
by using prolate spheroidal coordinates ($t,x,y,\phi$) with $x \geq 1$, $-1 \leq y \leq 1$;
the quantities $f$, $\omega$ and $\gamma$ are functions of $x$ and $y$ only and $\sigma$ is a constant.
The relation to Boyer-Lindquist coordinates $(t,r,\theta,\phi)$ is given by 
\beq
\label{trafBL}
t=t\ , \qquad
x=\frac{r-M}{\sigma}\ , \qquad 
y=\cos\theta\ , \qquad
\phi=\phi\ . 
\eeq

\subsection{Geodesics}

The geodesic motion of test particles is governed by the following equations \cite{bglq}
\begin{eqnarray}
\label{geoeqns}
\dot t&=&\frac{E}{f}+\frac{\omega f}{\sigma^2X^2Y^2}(L-\omega E)\ , \qquad 
\dot \phi=\frac{f}{\sigma^2X^2Y^2}(L-\omega E)\ , \nonumber\\
\ddot y&=&-\frac12\frac{Y^2}{X^2}\left[\frac{f_y}{f}-2\gamma_y+\frac{2y}{X^2+Y^2}\right]{\dot x}^2
+\left[\frac{f_x}{f}-2\gamma_x-\frac{2x}{X^2+Y^2}\right]{\dot x}{\dot y}\nonumber\\
&&+\frac12\left[\frac{f_y}{f}-2\gamma_y-\frac{2y}{X^2+Y^2}\frac{X^2}{Y^2}\right]{\dot y}^2\nonumber\\
&&-\frac12\frac{e^{-2\gamma}}{f\sigma^4X^2Y^2(X^2+Y^2)}
\left\{Y^2[f^2(L-\omega E)^2+E^2\sigma^2X^2Y^2]f_y\right.\nonumber\\
&&\left.+2(L-\omega E)f^3[y(L-\omega E)-EY^2\omega_y]\right\}\ , \nonumber\\
{\dot x}^2&=&-\frac{X^2}{Y^2}{\dot y}^2+\frac{e^{-2\gamma}X^2}{\sigma^2(X^2+Y^2)}\left[E^2-\mu^2f-\frac{f^2}{\sigma^2X^2Y^2}(L-\omega E)^2\right]\ , 
\end{eqnarray}
where Killing symmetries and the normalization condition $g_{\alpha\beta}\dot x^\alpha\dot x^\beta=-\mu^2$ have been used.
Here $E$ and $L$ are the conserved energy (associated with the Killing vector $\partial_t$) and angular momentum (associated with the Killing vector $\partial_\phi$) of the test particle respectively, $\mu$ is the particle mass and a dot denotes differentiation with respect to the affine parameter $\lambda$ along the curve; furthermore, the notation 
\beq
X=\sqrt{x^2-1}\ , \quad Y=\sqrt{1-y^2}\ 
\eeq
has been introduced.
For timelike geodesics, $\lambda$ can be identified with the proper time by setting $\mu=1$.
Let $U$ be the associated 4-velocity vector ($U\cdot U=-1$). 
Null geodesics are characterized instead by $\mu=0$.
Let $K$ be the associated tangent vector ($K\cdot K=0$).

Let us consider the motion on the symmetry plane $y=0$.
If $y=0$ and $\dot y=0$ initially, the third equation of Eqs. (\ref{geoeqns}) ensures that the motion will be confined on the symmetry plane, since the derivatives of the metric functions with respect to $y$, i.e., $f_y$, $\omega_y$ and $\gamma_y$, all vanish at $y=0$, so that $\ddot y=0$ too.
Eqs. (\ref{geoeqns}) thus reduce to
\begin{eqnarray}
\label{geoeqnsequat}
\dot t&=&\frac{E}{f}+\frac{\omega f}{\sigma^2X^2}(L-\omega E)\ , \qquad 
\dot \phi=\frac{f}{\sigma^2X^2}(L-\omega E)\ , \nonumber\\
{\dot x}&=&\pm\frac{e^{-\gamma}X}{\sigma\sqrt{1+X^2}}\left[E^2-\mu^2f-\frac{f^2}{\sigma^2X^2}(L-\omega E)^2\right]^{1/2}\ , 
\end{eqnarray}
where metric functions are meant to be evaluated at $y=0$.

\subsection{Neutrino oscillations}

The phase associated with neutrinos of different mass eigenstate is given by \cite{stodolsky}
\beq
\label{phase}
\Phi_k=\int_A^BP_{\mu\,(k)}\rmd x^\mu\ ,
\eeq
if the neutrino with 4-momentum $P=m_kU$ is produced at a spacetime point $A$ and detected at $B$.

The standard assumptions usually applied to evaluate the phase are the following (see, e.g., Ref. \cite{petcov}): 
a massless trajectory is assumed, which means that the neutrino travels along a null geodesic path;
the mass eigenstates are taken to be the energy eigenstates, with a common energy $E$; 
the ultrarelativistic approximation $m_k\ll E$ is performed throughout, so that all quantities are evaluated up to first order in the ratio $m_k/E$.

The integral is carried out over a null path, so that Eq. (\ref{phase}) can be also written as
\beq
\label{phase2}
\Phi_k=\int_{\lambda_A}^{\lambda_B}P_{\mu\,(k)}K^\mu\rmd \lambda\ , 
\eeq
where $K$ is a null vector tangent to the photon path. 
The components of $P$ and $K$ are thus obtained from Eq. (\ref{geoeqnsequat}) by setting $\mu=m_k$ and $\mu=0$ respectively.
In the case of equatorial motion the argument of the integral (\ref{phase2}) depends on the coordinate $x$ only, so that the integration over the affine parameter $\lambda$ can be switched over $x$ by  
\beq
\label{phase2equat}
\Phi_k=\int_{x_A}^{x_B}P_{\mu\,(k)}\frac{K^\mu}{K^x}\rmd x\ ,
\eeq
where $K^x=\rmd x/\rmd\lambda$.
By applying the relativistic condition $m_k\ll E$ we find 
\beq
\label{phase3}
\Phi_k\simeq\mp\frac12\sigma^2\frac{m_k^2}{E}\int_{x_A}^{x_B}\frac{xe^{\gamma}}{\sqrt{\sigma^2(x^2-1)-f^2(b-\omega)^2}}\rmd x\ , 
\eeq
to first order in the expansion parameter $m_k/E\ll1$, where $E$ is the energy for a massless neutrino and $b=L/E$ the impact parameter.

Therefore, the phase shift responsible for the oscillation is given by 
\beq
\label{shift}
\Phi_{kj}=\Phi_k-\Phi_j
\simeq\mp\frac12\sigma^2\frac{\Delta m_{kj}^2}{E}\int_{x_A}^{x_B}\frac{xe^{\gamma}}{\sqrt{\sigma^2(x^2-1)-f^2(b-\omega)^2}}\rmd x\ ,
\eeq
where 
\beq
\Delta m_{kj}^2=m_k^2-m_j^2\ .
\eeq

The question as to whether neutrino oscillations should be thought of as taking place between states of the same energy or the same momentum is still open.
The various controversies concerning quantum-mechanical derivations of the oscillation formula as well as the contradictions between the existing field-theoretical approaches proposed to settle them are reviewed in Ref. \cite{beuthe}.
The advantage of the equal-energy prescription is that the time dependence completely drops from the phase difference.
This is also justified by the fact that in none of the neutrino oscillation experiments the time was measured, only distances between creation and detection points, as discussed by Lipkin \cite{lipkin} and Stodolsky \cite{stodolsky2}.
They showed that in the plane wave approximation neutrino oscillations are experimentally observable only as a result of interference between neutrino states with different masses and the same energy. 
All interference effects between neutrino states having different energies are destroyed by the interaction between the incident neutrino and the neutrino detector \cite{lipkin2}.
The absence of clocks in these experiments allows to consider the behaviour of neutrinos as a \lq\lq stationary'' one, only the distance between the source and detector being known.
Therefore, the time interval is not an observable and only the oscillation wave length is measured.

When the separation between source and detector is large enough, the coherence between the different mass eigenstates is expected to be lost.
However, for atmospheric and solar neutrinos, where the source is free to move in distances many orders of magnitudes
larger, the decoherence distance will be even larger \cite{lipkin3}. This is in agreement with the result
quoted in Ref. \cite{kimbook} that the coherence is lost only at astronomical distances much larger than the size of the solar system and that this coherence loss is relevant only for supernova neutrinos.

\section{Neutrino oscillations in the Hartle-Thorne metric}

The exterior field of a slowly rotating slightly deformed object is described by the Hartle-Thorne metric \cite{ht67}, whose line element can be written in Lewis-Papapetrou form (\ref{metgen}) with metric functions
\begin{eqnarray}
f&\simeq&f_S\left[1-q\left(2P_2Q_2+\ln\frac{x-1}{x+1}\right)\right]-\frac{x^2+x-2y^2}{(x+1)^3}\left(\frac{a}{M}\right)^2\ , \nonumber\\
\omega&\simeq&2M\frac{1-y^2}{x-1}\left(\frac{a}{M}\right)\ , \nonumber\\
\gamma&\simeq&\gamma_S+2q(1-P_2)Q_1-\frac12\frac{1-y^2}{x^2-1}\left(\frac{a}{M}\right)^2\ ,
\end{eqnarray}
and $\sigma=\sqrt{M^2-a^2}$.
Here $P_l(y)$ and $Q_l(x)$ are Legendre polynomials of the first and second kind respectively, while the functions
\beq
\label{schw}
f_S=\frac{x-1}{x+1}\ , \qquad
\gamma_S=\frac12\ln\left(\frac{x^2-1}{x^2-y^2}\right)
\eeq
correspond to the Schwarzschild solution ($q=0=a$). 
The connection between this form of the metric and the original one as derived by Hartle and Thorne was discussed in Ref. \cite{bglq}.

When expressed in terms of the standard Boyer-Lindquist coordinates (\ref{trafBL}) the phase shift (\ref{shift}) becomes
\begin{eqnarray}
\label{shifthtBL}
\Phi_{kj}&\simeq&\mp\frac{\Delta m_{kj}^2}{2E}
\left[r-\frac{M^2}{r}\left(q+\frac{b^2-a^2}{2M^2}\right)-\frac{M}{2r^2}\left[q(b^2+2M^2)-(a+b)^2\right]\right]_{r_A}^{r_B}\nonumber\\
&=&\mp\frac{\Delta m_{kj}^2}{2E}(r_B-r_A)
\left[1+\frac{M^2}{r_Br_A}\left(q+\frac{b^2-a^2}{2M^2}\right)\right.\nonumber\\
&&\left.+\frac{M(r_B+r_A)}{2r_B^2r_A^2}\left[q(b^2+2M^2)-(a+b)^2\right]\right]\ ,
\end{eqnarray}
where terms of the order of $q(a/M)$ as well as higher order terms in the weak field expansion $M/r\ll1$ have been neglected.
In the limiting case of vanishing quadrupole parameter ($q=0$) Eq. (\ref{shifthtBL}) reproduces the results of Ref. \cite{kerr} for the slowly rotating Kerr spacetime.
It is also useful to replace the parameters $a/M$ and $q$ by the Hartle-Thorne angular momentum ${\mathcal J}$ and mass quadrupole moment ${\mathcal Q}$ according to
\beq
{\mathcal M}=M(1-q)\,,\qquad
{\mathcal J}=-Ma\,,\qquad
{\mathcal Q}=Ma^2+\frac{4}{5}M^3q\,.
\eeq
For instance, for $b=0$ Eq. (\ref{shifthtBL}) gives
\begin{eqnarray}
\label{shifthtBL2}
\Phi_{kj}&\simeq&\mp\frac{\Delta m_{kj}^2}{2E}(r_B-r_A)
\left[1+\frac{{\mathcal M}^2}{r_Br_A}\left(\frac{5}{4}\frac{{\mathcal Q}-7{\mathcal J}^2/{\mathcal M}}{{\mathcal M}^3}\right)\left(1+\frac{{\mathcal M}(r_B+r_A)}{r_Br_A}\right)\right] \nonumber\\
&\equiv&\Phi_{kj}^{\rm(mono)}+\Phi_{kj}^{\rm(dip)}+\Phi_{kj}^{\rm(quad)}\ ,
\end{eqnarray}
where
\begin{eqnarray}
\label{shifthtBL3}
\Phi_{kj}^{\rm(mono)}&=&\mp\frac{\Delta m_{kj}^2}{2E}(r_B-r_A)\ , \nonumber\\
\Phi_{kj}^{\rm(dip)}&=&\Phi_{kj}^{\rm(mono)}\frac{{\mathcal M}^2}{r_Br_A}\left(1+\frac{{\mathcal M}(r_B+r_A)}{r_Br_A}\right)\left(-\frac{35}{4}\frac{{\mathcal J}^2}{{\mathcal M}^4}\right)\ , \nonumber\\
\Phi_{kj}^{\rm(quad)}&=&\Phi_{kj}^{\rm(mono)}\frac{{\mathcal M}^2}{r_Br_A}\left(1+\frac{{\mathcal M}(r_B+r_A)}{r_Br_A}\right)\left(\frac{5}{4}\frac{{\mathcal Q}}{{\mathcal M}^3}\right)\ .
\end{eqnarray}
The monopole term is the dominant one, due to the large distance between source and detector. 
However, describing the background gravitational field simply by using the spherically symmetric Schwarzschild solution is not satisfactory in most situations.
In fact, astrophysical sources are expected to be rotating as well endowed with shape deformations leading to effects which cannot be neglected in general.
The modification to the phase shift induced by spacetime rotation has been already taken into account in Ref. \cite{kerr}.
We will estimate below the contribution due to the quadrupole moment of the source with respect to that due to spin for solar, atmospheric and astrophysical neutrinos by evaluating the ratio 
\beq
\frac{\Phi_{kj}^{\rm(quad)}}{\Phi_{kj}^{\rm(dip)}}=-\frac17\left(\frac{{\mathcal Q}}{{\mathcal M}^3}\right)\left(\frac{{\mathcal M}^4}{{\mathcal J}^2}\right)\ .
\eeq

In the case of the Sun we have ${\mathcal M}_\odot\approx1.5\times10^{5}$ cm,  $R_\odot\approx7\times10^{10}$ cm and
${\mathcal J}_\odot\approx5\times10^{9}$ cm$^2$, so that
\beq
\frac{{\mathcal M}_\odot}{R_\odot}\approx2\times10^{-6}\,, \qquad
\frac{{\mathcal J}_\odot}{R_\odot^2}\approx10^{-12}\,, \qquad
\frac{{\mathcal Q}_\odot}{R_\odot^3}\approx-4\times10^{-13}\,, 
\eeq
where the mass quadrupole moment has been evaluated through the relation ${\mathcal Q}_\odot=-J_2{\mathcal M}_\odot R_\odot^2$, with $J_2\approx2\times10^{-7}$ (see, e.g., Ref. \cite{rozelot}).
Therefore, $[{\mathcal J}_\odot^2/{\mathcal M}_\odot]/{\mathcal M}_\odot^3\approx5\times10^{-2}$ and ${\mathcal Q}_\odot/{\mathcal M}_\odot^3\approx-4.4\times10^{4}$, so that the quadrupole contribution is $10^5$ times greater than that due to spin.

In the case of the Earth the quadrupole term is even more dominant.
In fact, since ${\mathcal M}_\oplus\approx0.45$ cm,  $R_\oplus\approx6.4\times10^{8}$ cm and
${\mathcal J}_\oplus\approx1.5\times10^{2}$ cm$^2$, we have
\beq
\frac{{\mathcal M}_\oplus}{R_\oplus}\approx7\times10^{-10}\,, \qquad
\frac{{\mathcal J}_\oplus}{R_\oplus^2}\approx3.7\times10^{-16}\,, \qquad
\frac{{\mathcal Q}_\oplus}{R_\oplus^3}\approx-7\times10^{-13}\,, 
\eeq
with ${\mathcal Q}_\oplus=-J_2{\mathcal M}_\oplus R_\oplus^2$ and $J_2\approx10^{-3}$, implying that $[{\mathcal J}_\oplus^2/{\mathcal M}_\oplus]/{\mathcal M}_\oplus^3\approx5.5\times10^{5}$ and ${\mathcal Q}_\oplus/{\mathcal M}_\oplus^3\approx-2\times10^{15}$ leading to $\Phi_{kj}^{\rm(quad)}/\Phi_{kj}^{\rm(dip)}\approx5\times10^8$.

The contribution due to rotation is expected to be comparable with that due to the deformation in the case of rotating neutron stars. 
Laarakkers and Poisson \cite{poisson} numerically compute the mass quadrupole moment ${\mathcal Q}_{\rm NS}$  for several equations of state.
They found that for fixed gravitational mass ${\mathcal M}_{\rm NS}$, the quadrupole moment is given as a simple quadratic fit, i.e., 
\beq
{\mathcal Q}_{\rm NS}\simeq-k\frac{{\mathcal J}_{\rm NS}^2}{{\mathcal M}_{\rm NS}}\,,
\eeq
where ${\mathcal J}_{\rm NS}$ is the angular momentum of the star and $k$ is a dimensionless parameter which depends on the equation of state. It varies between $k\sim2$ for very soft equations of state and $k\sim8$ for very stiff ones, for a typical mass of ${\mathcal M}_{\rm NS}=1.4M_\odot$. 
Therefore, in this case the spin term in Eq. (\ref{shifthtBL2}) is of the same order as the quadrupole term.

Finally, it is worth to note that the monopole term in Eq. (\ref{shifthtBL2}) is always the leading one.
In fact, for solar neutrinos the factor ${\mathcal M}_\odot^2/r_Br_A$ turns out to be about $2.3\times10^{-12}$, assuming $r_A\sim r_\oplus$ and $r_B\sim R_\oplus+d\sim d$, being $d\approx1.5\times10^{13}$ cm the Earth-Sun distance.
The first correction to the monopole term is thus of the order of $10^{-7}$. 
For atmospheric neutrinos ${\mathcal M}_\oplus^2/r_Br_A\sim{\mathcal M}_\oplus^2/R_\oplus^2\approx5\times10^{-19}$, implying that the first correction is of the order of $10^{-3}$.

\section{Concluding remarks}

The issue of the interaction of neutrinos with gravitational fields is timely and has a lot of implications in astrophysics and cosmology.
Neutrinos are usually generated by radioactive decays or nuclear reactions such as those occurring in the Sun, stars, accelerators or nuclear reactors and in particular when cosmic rays hit atoms.
Neutrinos are also expected to carry off the largest amount of energy of an exploding star in a supernova.
Neutrinos from core collapse supernovae can be emitted from a rapidly accreting disk surrounding a black hole and have been suggested to be responsible for the cooling process \cite{popham}.
In general the coalescence of compact objects produces hot disks. 
The neutrino flux from these objects is so large that it would be easily detected by currently online neutrino detectors \cite{mcLaughlin}.

Neutrino detectors of increasing sensitivity require a better understanding of propagation and oscillation properties of neutrinos in the neighborhood of massive astrophysical objects.
In particular, it has been suggested that the gravitational oscillation phase might have a significant effect in supernova explosions \cite{grossman}.

We have evaluated the correction to the phase difference of neutrino mass eigenstates due to the gravitational field produced by a massive, slowly rotating and quasi-spherical object, described by the Hartle-Thorne metric, generalizing previous results.
This is an approximate solution of the vacuum Einstein equations accurate to second order in the rotation parameter and to first order in the mass quadrupole moment of the source, generalizing the Lense-Thirring metric.
The large distances covered by neutrinos from the emitting source to the detector allow to apply the weak field limit in the calculations, so that the Hartle-Thorne metric is enough to account for the leading gravitational effects.     
We have shown that, apart from the monopole term which is the dominant one, the contribution due to the quadrupole moment is much greater than that due to spin in the case of solar neutrinos and even more for atmospheric neutrinos.
The effect of rotation is instead expected to be comparable with that due to deformation in the case of rotating neutron stars. 

Our result can also be relevant in the context of CPT and lepton number violation in the neutrino sector induced by gravity (see, e.g., Ref. \cite{sinha} and references therein).
In fact, it has been recently argued that the generation of a neutrino asymmetry can arise in accretion disks around rotating black holes or more generally when neutrinos are propagating in background spacetimes which are not spherically symmetric \cite{mukhopdhyay}. 
The proposed CPT violation mechanism leading to neutrino oscillations would be due to the different spin-gravity coupling of neutrinos with respect to anti-neutrinos.
Of course, this is only one of all possible scenarios. 
Observable neutrino oscillations may result from a combination of effects involving neutrino masses and Lorentz violation \cite{coleman}.
Similar effects can also arise from violations of the equivalence principle \cite{gasperini}.
Both current and future experiments on neutrino oscillations are expected to clarify the interplay of gravity and neutrino physics.

\section*{Acknowledgements}

The authors thank ICRANet for support.
OL acknowledges Profs. H. Quevedo and S. Capozziello for useful discussions.

\end{document}